# Evaluating Different Cost-Benefit Analysis Methods for Port Security Operations

Galina Sherman (a)*, Peer-Olaf Siebers (b), David Menachof (a), Uwe Aickelin (b)

(a) Business School, Hull University, Hull HU6 7RX, UK
(b) School of Computer Science, The University of Nottingham, Nottingham NG8 1BB, UK
*G.Sherman@2008.hull.ac.uk

**Abstract**

Service industries, such as ports, are attentive to their standards, a smooth service flow and economic viability. Cost benefit analysis has proven itself as a useful tool to support this type of decision making; it has been used by businesses and governmental agencies for many years. In this book chapter we demonstrate different modelling methods that are used for estimating input factors required for conducting cost benefit analysis based on a single case study. These methods are: scenario analysis, decision trees, Monte-Carlo simulation modelling and discrete event simulation modelling. Our aims are, on the one hand, to guide the analyst through the modelling processes and, on the other hand, to demonstrate what additional decision support information can be obtained from applying each of these modelling methods.

## 1. Introduction

Businesses are interested in the trade-off between the cost of risk mitigation and the expected losses of disruptions (Kleindorfer and Saad 2005). Port operations, due to their complexity, have a large impact on the economy, e.g. a paralysed port will have an enormous effect on an area or country. Airports and seaports have an additional complexity when conducting such risk analysis, because there are two key stake holders with different interests involved in the decision processes concerning the port operation: Port operators and security operators (Bichou 2004). In this chapter we focus on the tangible costs and benefits for port stake holders.

Airports and sea ports as well as other service industries are interested in smooth traffic flows, short service times at different stations, quality and efficiency. In ports we have port operators which are service providers and as such interested in all the performance parameters mentioned above and we have the border agency which represent national security interests that need to be considered. However, the interests of the border agency are different from ones of the service providers. Detection of threats such as weapons, smuggling and sometimes even stowaways is the high importance. Nevertheless, security checks that are too long can compromise the service standard targets that port operators are trying to maintain.

Besides these two conflicting interest there is also the cost factor for security that needs to be kept in mind. Security checks require expensive equipment and well trained staff. However, the consequences for the public of undetected threats crossing the border can be severe. Therefore, it is in the interest of all parties to find the right balance between service, security, and costs.

How can we decide the level of security required to guarantee a certain detection threshold for threats while maintaining economic viability and avoiding severe disruptions to the process flow? A tool frequently used by business and government officials to support such investigations is Cost-Benefit Analysis (CBA) (Hanley and Spash 1993). Compared to other methods such as Cost Effectiveness Analysis (CEA) or Cost Utility Analysis (CUA) it provides a monetary value quantifying the expected benefits and costs. In order to use CBA, certain input parameters need to be known or, in absence of real data, estimated.

Methods frequently used to estimate these input parameters are Scenario Analysis (SA), Decision Trees (DT) and Monte Carlo Simulation (MCS) (Damodaran 2007). Discrete Event Simulation (DES) is less frequently used in risk analysis but often used in Operational Research (Turner and Williams 2005; Wilson 2005) and, in particular, in service industries (Laughery et al. 1998) (e.g. banks, medical, transportation) to investigate different operational practices. When reviewing the relevant literature on risk analysis there appears to be a gap comparing the efficiency of all these methods using a single case study (Virta et al. 2003, Jacobson et al. 2006).

In this book chapter, we present such a comparison of different probabilistic techniques that are used for conducting CBA. In particular, we focus on SA, DT and simulation. We demonstrate step-by-step the application of these methods, while conducting a CBA of different cargo screening policies. Our main aim is to demonstrate how CBA can be applied to service industry problems. Furthermore, we want to show what the data requirements are for the different modelling methods and what additional decision support intelligence an analyst can obtain by applying each of the modelling methods mentioned above. This information is intended to support the analyst in the decision making process of which method to use for their analysis.

Section 2 comprises a brief review of the existing literature. First, we focus on the advantages and disadvantages of CBA compared to alternative methods. Then we look at case study examples from the service industry where the different modelling methods mentioned above have been applied. In Section 3, we introduce our case study (cargo screening process) and show step-by-step how to conduct a CBA and how to apply the different modelling methods to the case study scenario. In Section 4, we summarise our finding and conclude.

## 2. Literature Review

*2.1 CBA and alternative analysis methods*

The use of CBA is described in the literature from the early 19th century. Already at that time this approach was used by US governmental agencies in environmental management (Hanley and Spash 1993). According to Guess and Farnham (2000), since 1960 the application of CBA was expanded to "human beings" and "physical investment programs".

There are some alternatives to CBA such as CEA and CUA. The economics literature compares these three methods. The difference between these three approaches is that CBA allows a comparison of wider range of scenarios, because the costs and the results have a monetary expression unlike two other techniques that are observed by using a single result every time. Thus the advantage of CBA over the other two methods is that it has wider scope of possibilities and is more relevant for the service sector. However, sometimes it is impossible to give a monetary value to the costs or to the benefits (Guess and Farnham 2000).

A related method is Risk Analysis (RA). "RA consists of the repeated random extraction of a set of values for the critical variables, taken within the respective defined intervals, and then calculating the performance indices for the project resulting from each set of extracted values". It does not help to reduce the risk itself, but to identify and manage it (EC 2008). Our work has communalities with both RA and CBA, and in fact is a form of CBA that assesses risks.

A competitive method to CBA that we find in the literature is Multi Criteria Analysis (MCA). Our method allows a comparison between a mixture of inputs- monetary and non-monetary. It can use the results of a standard CBA as monetary inputs and service quality estimators as non-monetary inputs. From these it produces results to show the relationship between costs and benefits of different options (DCLG 2009). In this chapter we conduct a relatively simple CBA study based on a single case of port operation, this can be extended in the future using MCA.

De Langen and Pallis (2006) conduct a qualitative study to analyse the benefits of existence of the intra port competition. We learn that the use of costs vs. benefits can be valuable at many aspects of the port performance. In their research the authors use a qualitative methodology (e.g. survey), which we find less beneficial for our paper. Jacobson et al. (2006) and Virta et al. (2003) suggest that CBA in evaluation of trade offs and cost effectiveness while screening 100% of the cargo and using single or double screening devices. The authors suggest a cost model that contains direct costs and indirect costs associated with costs of failure.

Bichou (2004) finds that CBA is a model used to evaluate optimal policy decisions. According to him this model is useful while port authorities are interested to add value, e.g. a programme which is non mandatory as long as the benefits of the new behaviour can be quantified. According to Bichou (2004) there are, however, some difficulties to apply cost analysis to port performance and one of them is the difficulty to estimate efficiency measures. Moreover, Bichou et al. (2009) argue that an additional concern in using a CBA while dealing with security issues arises as a result of uncertainty, which means that the benefits will be also uncertain or difficult to be quantified and evaluated, e.g. in case of a terrorist threat. They find that it is very difficult to judge the effectiveness of the security policy. On the other hand they argue that an additional difficulty lies in the fact that one cannot know what would happen if the policy would not exist.

While we acknowledge Bichou's et al. (2009) argument that conducting a CBA can be limited when dealing with rare security issues such as terrorist attacks. We tend to agree with other authors such as Jacobson et al. (2006) and Virta et al. (2003) who suggest using CBA to compare between different screening policies for better service and performance routines.

In conclusion, we find that CBA is a powerful tool for comparing costs vs. benefits of different practices in service sector. It allows to a user to compare a wide numbers of variables and provides a monetary value to the comparison. We find CBA suitable for our research purpose.

*2.2 Modelling methods applied in Service Industry CBA*

*2.2.1 Scenario analysis*

SA is a version of sensitivity analysis which is used to study the possible impact on the outcomes, while different variables are tested (EC 2008). It is often used to analyse possible future scenarios by considering possible best, worst, and average outcomes. According to Damodaran (2007), this technique is suitable for single events. Daellenbach and McNickle (2005), state that any business faces uncertainty and as a result creates an unlimited number of possible futures to be considered. However, the number of possible scenarios to be considered is limited to three or four in scenario analysis. Bruzzone et al. (1999) use this method to verify the suitability of the simulators of container terminals.

*2.2.2 Decision trees*

A DT is a decision support tool (diagram) used in operational research. It can be helpful in deciding about strategies and dealing with conditional probabilities. According to Anderson et al. (1985), DTs are particularly useful when dealing with relatively few possible solutions. DTs are diagrams that can be used to represent decision problems so that their structure is made clearer. Unlike decision tables, DT can be used to represent problems involving sequences of decisions, where decisions are made at different stages in the problem. For instance, Kim et al. (2000) use DTs to decide about storage policies for transshipments.

*2.2.3 Simulation*

Simulation is widely used in the areas of logistics. Turner and Williams (2005) specify that simulating the behaviour of complicated systems gives the ability to experiment with different scenarios and has the power of generalisation of the insight on the performance of the complicated systems. Also DES is often used in modelling of traffic movement (Robinson 2004, Laughery et al. 1998) and traffic and transport management (Davidsson 2005). Furthermore, Wilson (2005) confirms the usefulness of simulation for investigating security issues and states that "Simulation modeling allows the analysis or prediction of operational effectiveness, efficiency, and detection rates (performance) of existing or proposed security systems under different configurations or operating policies before the existing systems are actually changed or a new system is built, eliminating the risk of unforeseen bottlenecks, under- or over-utilization of resources, or failure to meet specified security system requirements".

In the medical literature, we find a successful use of simulation for CBA. For example, Habbema et al. (1987) suggest simulation as an appropriate technique to conduct the CBA to compare two different cancer screening policies. The authors use a micro simulation approach to explore different scenarios and their outputs. Also Pilgrim et al. (2009) suggest conducting a CEA for cancer screening policies while using discrete event simulation. The authors support their choice of methodology with previous research using the same research strategy.

As we can learn from the literature above simulation is widely used in different service industries e.g. medical, transportation, etc. (Laughery et al. 1998). It is a powerful tool that can help to an analyst to learn about the system under research. In addition simulation allows the user to compare between different scenarios before implementing them in real world.

There are different types of simulation and here we focus on static discrete stochastic simulation (i.e. MCS) and dynamic discrete stochastic simulation (i.e. DES). The difference between them is that in MCS time does not play a natural role, but in DES it does (Kelton et al., 2010). While the first is often used in risk assessment the second is used when further investigation into the system behaviour on the operational level are required for the decision making.

## 3. Case study

In this section we apply the approaches described above in a real world situation. Furthermore, Farrow and Shapiro (2009) identify that the literature dealing with the CBA is often based on a case study methodology while focusing on a sensitive or strategic topic. We find this approach applicable for the service industry as well.

### *3.1 The Case Study System*

In our research we chose to make the comparison of SA, DT and simulation for a CBA based on the same real world system, which is different from the practices we find in the literature where we find different example for each approach (Damodaran 2007). We find that the users can benefit and learn more if a comparison of methods is conducted, based on the same data. We chose a case study approach as our research methodology it will supply us the data needed for all the methods. Our case study involves the cargo screening facilities of the ferry port of Calais (France). This case study was conducted in collaboration with the UK Border Agency (UKBA).

At Calais we find two main stake holders related to service provision: port operators which are service providers and as such interested in a smooth flow of port operations to provide certain service standards (e.g. service times) and the border agency which represent national security interests that need to be considered. Checks have to be conducted to detect threats such as weapons, smuggling and stowaways. If the security checks take too long they can compromise the service standard targets to be achieved by the port operators. We have chosen Calais as our case study system for two reasons: First, it has limited number of links – Calais operates only with Dover leading to a simple cargo flow. Second, there is only one major threat of interest to the British government - clandestines. These are people who are trying to enter the UK illegally – i.e. without having proper papers and documents.

In Calais there are two security areas, one is operated by French authorities and one is operated by UKBA. According to the data collected between April 2007 and April 2008 about 900,000 lorries passed the border and approximately 0.4% of these lorries have been found to have an additional human freight (UKBA 2008). More details can be found in Table 1.

Table 1: Statistics from Calais (from April 2007 - April 2008)

| Statistic | Value |
| --- | --- |
| Total number of lorries entering Calais harbour | 900000 |
| Total number of positive lorries found | 3474 |
| Total number of positive lorries screened on French site | 900000 |
| Total number of positive lorries found on French site | 1800 |
| Total number of positive lorries screened on UK site | 296406 |
| Total number of positive lorries found on UK site | 1674 |
| … In UK Sheds | 890 |
| … In UK Berth | 784 |

The search for clandestines is organised in three major steps, one by France and two by the UKBA. On the French side, after passing the passport check (referred to as Passport in the DT) all arriving lorries are screened, using passive millimetre wave scanners (PMMW) for soft sided lorries and heartbeat detectors (HB) for hard sided lorries. If lorries are classified as suspicious after the screening further investigations are undertaken. For soft sided lorries there is a second test with $CO_2$ probes (CO2) and if the result is positive the respective lorry is opened. For hard sided lorries there is no second test and they are opened immediately. The cleared lorries proceed to purchase a ticket for the ferry and then to the UK passport check and screening.

On the British side only a certain percentage of lorries (currently 33%) is searched at the British sheds. Here a mixture of measures is used for the inspection, e.g. $CO_2$ probes, dogs and opening lorries. Once the lorries passed the British sheds they will park in the berth to wait for the ferry. In the berth there are mobile units operating that search as many of the parked lorries as possible before the ferry arrives, using the same mixture of measures than in the sheds. As shown in Table 1 only about 50% of the clandestines detected were found by the French, about 30% in the sheds and 20% by the mobile units in the berth. The overall number of clandestines that are not found by the authorities is of course unknown.

*3.2 Analysis framework*

We consider two factors with three scenarios each in our analysis: Traffic Growth (TG) and Clandestine Growth (CG) (Table 2). For each factor and scenario combination, we have estimated the probability of it happening, as described in the following paragraphs. The question we are trying to answer is how the UKBA should respond to these scenarios. We assume that there are three possible responses: increasing the searches by either 0%, 10% or 20%.

Our TG scenarios are based on estimates by the port authorities who are planning to build a new terminal in Calais in 2020 to cope with all the additional traffic expected. According to DHB (2008) between 2010 and 2020 the traffic in the Port of Dover is expected to double. Due to their direct connection, one can assume that this is also applicable to the Port of Calais. Thus an annual traffic growth of 0%-20% is a realistic factor range, with an increase of 10% most likely, while the other two are equally likely. It is assumed that any increase in traffic is proportional, i.e. the ratio of soft to hard sided lorries remains the same. It is an important assumption because different types of lorries are screened by different technologies. If the ratio between them changes it might cause to the bottlenecks at some places and empty and waiting staff at others.

The second factor under consideration is CG. This is the most unpredictable of the three factors, as clandestine numbers greatly vary from year to year based largely on external factors such as the economic attractiveness of the UK, the number and intensity of wars and other conflicts worldwide and other political initiatives.

Local aspects also play a role, e.g. an increase in searches in Calais can displace clandestines to other nearby ports and vice versa. Due to the uncertainty attached to this factor, a range of +25% to -50% is considered, with all scenarios being equally likely. A higher maximum decrease than increase is assumed to the recent clearing in late 2009 of the Calais "jungle" (illegal encampment of clandestines near the port). We will assume in the following that any changes in clandestine numbers will proportionally affect successful and unsuccessful clandestines.

Search Growth (SG) describes the percentage increase in search activity by the UKBA. Currently, UKBA searches 33% of traffic. To keep this proportion stable, UKBA will need to respond to a growth in traffic by increasing the number of lorries it searches. At the same time, there is political pressure to search more vehicles, whilst budget pressures limit the number of vehicles that can be inspected. Searching 33% of the traffic represents a trade-off between the service levels and the investments. Thus we assume that search growth may also vary between 0% and 20% and not higher because of costs. As before, we assume that any increase in search activity is proportional to hard and soft sided lorries.

Table 2: two factors with three scenarios and one decision variable with three options.

| Traffic Growth (TG) | p(TG) | Clandestine Growth (CG) | p(CG) | Search Growth (SG) |
|---|---|---|---|---|
| +0% | 0.25 | -50% | 0.33 | +0% |
| +10% | 0.5 | +0% | 0.33 | +10% |
| +20% | 0.25 | +25% | 0.33 | +20% |

Combining the above information, we arrive at the following combined probabilities of each scenario to occur p(TG,CG)=p(TG)*p(CG) the results can be seen in Table 3.

Table 3: combined probabilities assuming independence of probabilities

|  | -50% CG | 0% CG | +25% CG |
|---|---|---|---|
| 0% TG | 0.083 | 0.083 | 0.083 |
| 10% TG | 0.167 | 0.167 | 0.167 |
| 20% TG | 0.083 | 0.083 | 0.083 |

It is estimated by the UKBA that each clandestine that reaches the UK costs the government approximately £20,000 per year. Moreover, it is estimated that the average duration of a stay of a clandestine in the UK is five years, so the total cost of each clandestine slipping through the search in Calais is £100,000. In addition, the average number of clandestines on a lorry is four, which means this cost £400,000 per positive lorry that is missed.

The cost for increasing the search capacity in Calais is more difficult to estimate, as there is a mixture of fixed and variable cost and operations are often jointly performed by French, British and private contractors. However, if we concentrate on UKBA's costs, we can arrive at some reasonable estimates, if we assume that any increase in searches would result in a percentage increase in staff and infrastructure cost. Thus we estimate that a 10% increase in search activity (10% SG) would cost £5M and a 20% increase £10M (20% SG) (Table 4).

Table 4: Cost of extra searches – as mentioned before.

| TG vs. SG | SG 0% | SG +10% | SG +20% |
| --- | --- | --- | --- |
| TG 0% | £0 | £5,000,000 | £10,000,000 |
| TG 10% | £0 | £5,000,000 | £10,000,000 |
| TG 20% | £0 | £5,000,000 | £10,000,000 |

As 33% of vehicles are searched by UKBA, we can calculate resultant percentages of vehicles searched by combining the above two factors (assuming linear relationships). For example, if traffic increase is matched by search increase, this will remain the same. Or if there is a +10% TG and +0% SG results in 33% * (100%/110%) = 30% of vehicles searched (Table 5).

A key question is the relationship between the percentage of vehicles searched vs. the number of clandestines found or more importantly, the number of clandestines not found. Searching 33%, UKBA finds approximately 1,674 lorries in Calais with additional cargo. A best estimate of "successful" clandestines is approximately 50 per month (600 per year) or 150 lorries per year. Establishing a clear relationship between these 150 and the figure of 1,674 is difficult, as 1,674 lorries do not represent unique attempts by the clandestines. Unsuccessful clandestines will try again, i.e. finding one more positive lorry does not remove four clandestines from the system.

Based on advice by the UKBA, we estimate the effect of increased searches as follows: a search increase of 10% yields an extra 167 positive lorries found. 10% more positive lorries detected is estimated to reduce the number of successful clandestines by 10%. Thus the number of non-detected positive lorries reduces from 150 lorries by 15 to 135 lorries. Please note that these numbers vary for the more detailed simulation scenarios due to effects such as queue jumping, i.e. a 10% increase in searches will yield less than a 10% reduction in undetected positive lorries.

It is probably a fair assumption that an increase in searches will yield a decrease in the number of successful clandestines and vice versa. In absence of further information and considering that the variation of percentage of searches is in a relatively limited range of 27.5% to 39.6%, we will make the same assumption here as in the rest of the scenario analysis: the relationship between both parameters is linear. Based on this, we obtain the number of clandestines missed as given in Table 5, e.g. searching only 30% of traffic results in 600* (33%/30%) = 660 missed clandestines.

Table 5: Proportion of vehicles searched

| TG vs. SG | SG 0% | SG +10% | SG +20% |
|---|---|---|---|
| TG 0% | 0.3300 | 0.3630 | 0.3960 |
| TG 10% | 0.3000 | 0.3300 | 0.3600 |
| TG 20% | 0.2750 | 0.3025 | 0.3300 |

For our case study system we conduct a SA, build a DT that fully represents the traffic flow inside the system boundaries, and build a simulation model of the system using the same input data as for the decision tree. Comparing the results of all these methods allowed us to validate the models. While we use Microsoft Excel for SA and modelling DT, our simulation models are implemented in AnyLogic (XJ Technologies 2010), a multi-paradigm simulation package.

We use different methods for estimating the 'adjusted' number of positive lorries found if there is no growth of positive lorries. Once we have the matrix (iteration over our two factors) we conduct some data analysis to estimate the costs. The data analysis is the same for all the different approaches we will present. We demonstrate this for SA and for all others we will only report on the key outputs (adjusted number of positive lorries found for CG=0, total expected costs). Finally, we will compare the expected costs that resulted from the different models and demonstrate the inputs and outputs of each one of the tools.

*3.3 Scenario analysis*

Table 6 is the result from the SA: number of positive lorries found. The cost estimation that follows below is the same for all methods discussed in this book chapter and will only be shown once in detail.

Table 6: Adjusted number of positive lorries found if CG = 0%. Calculated based on Table 5 where 0.3300 represents 1,674 lorries.

| TG vs. SG | SG 0%  | SG +10% | SG +20% |
|-----------|--------|---------|---------|
| TG 0%     | 1674.0 | 1841.4  | 2008.8  |
| TG 10%    | 1521.8 | 1674.0  | 1826.2  |
| TG 20%    | 1395.0 | 1534.5  | 1674.0  |

From Table 6 we can learn that if the TG 20% but the SG stays the same the number of positive lorries detected will be 1395, and other way around if the traffic does not grow (TG 0%) but the SG 20% the number of detected lorries will grow as well.

Table 7: Relative number of positive lorries found when compared to base scenario if CG = 0%

| TG vs. SG | SG 0% | SG +10% | SG +20% |
|---|---|---|---|
| TG 0% | 1 | 1.1 | 1.2 |
| TG 10% | 0.909091 | 1 | 1.090909 |
| TG 20% | 0.833333 | 0.916667 | 1 |

From Table 7 we can learn that comparing the number of positive lorries from Table 6 to the factor of CG = 0% the 1395 represents 83.3333% of the positive detected lorries.

Table 8: Number of positive lorries missed if CG = 0%. Calculated based on the probabilities in Table 5, where 0.3300 results in 150 missed positive lorries. Similar Tables can be computed for the other CG values.

| TG vs. SG | SG 0% | SG +10% | SG +20% |
|---|---|---|---|
| TG 0% | 150.0 | 136.4 | 125.0 |
| TG 10% | 165.0 | 150.0 | 137.5 |
| TG 20% | 180.0 | 163.6 | 150.0 |

Table 9: Relative number of positive lorries missed compared to the base scenario (1 means 150 lorries)

| TG vs. SG | SG 0% | SG +10% | SG +20% |
|---|---|---|---|
| TG 0% | 1.00 | 0.91 | 0.83 |
| TG 10% | 1.10 | 1.00 | 0.92 |
| TG 20% | 1.20 | 1.09 | 1.00 |

Table 8 presents the number of positive lorries missed for variable CG=0%. Similar results were calculated for CG=25% and CG=-50%. Table 9 represents an optional scenario number of positive lorries missed (e.g. 180) divided by base scenario (150 positive lorries missed) for TG 20%. From here we can learn that if the traffic will grow at 20% but search percentage will not grow, an extra of 20% of the positive lorries will be missed. Table 10 was calculated by using ratios from Table 9 multiplied by number of positive lorries missed (150 - not detected) multiplied by cost per missed positive lorry (£400,000) plus cost of extra searches (Table 4).

Table 10: Expected costs including SG costs for CG = 0%. Calculated by combining the information of Table 4 with that of Table 9, where cost of one positive lorry is £400,000 each.

| TG vs. SG | SG 0% | SG +10% | SG +20% |
|---|---|---|---|
| TG 0% | £60,000,000 | £59,545,455 | £60,000,000 |
| TG 10% | £66,000,000 | £65,000,000 | £65,000,000 |
| TG 20% | £72,000,000 | £70,454,545 | £70,000,000 |

Table 10 can be used to calculate the tables for CG=-50% and CG=25% by multiplying each value in the table with 1+CG(x) where CG(x) is the proportion of CG. The expected costs for each policy according to SA are as follows from Table 11. It was calculated by multiplying the probabilities of each scenario to occur with its costs e.g. $C(SG)=\sum(EC(SG,TG,CG)*p(TG,CG))$.

Table 11: Expected costs

| SG=0% | SG=10% | SG=20% |
|---|---|---|
| £60,500,000 | £60,000,000 | £60,416,667 |

Overall, we can conclude from the SA that the best option seems to be to change SG by 10%. In following DT and simulation we use the same assumptions, costs and scenarios as described above. When we use additional data for an approach we will state it in the relevant section.

*3.4 Decision tree*

In order to conduct a comparison between different techniques we used for the DT the same scenarios that we have used for the SA. Also, here we considered the same two factors: TG and CG and as a reaction SG. The aim is to compare the inputs and outputs of different tools and the level of details needed; also to investigate what is an appropriate policy that UKBA should adopt according to the DT. For our case study we have built a DT that fully represents the flow inside our system (see Figure 1). On basis of the DT we also built a MCS. Building both allows us to check the results and validate the models as the results should be identical.

Figure 1: Decision Tree

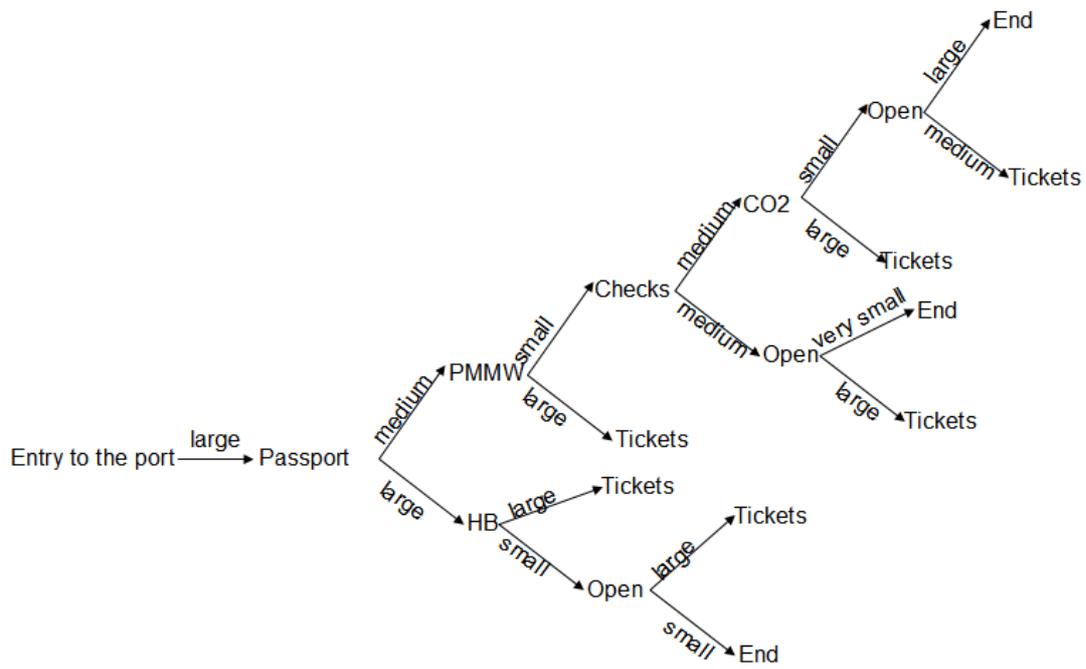

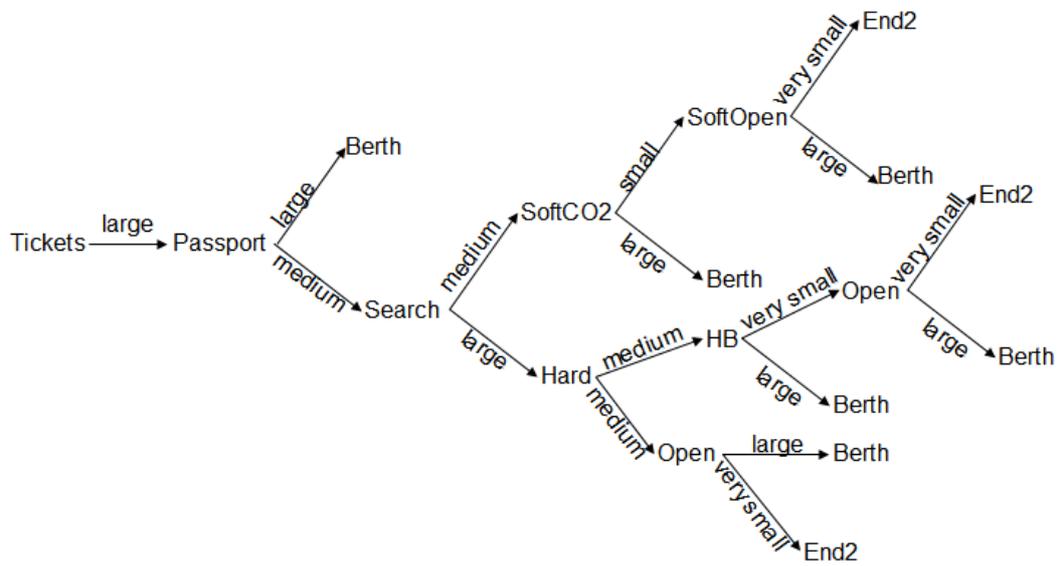

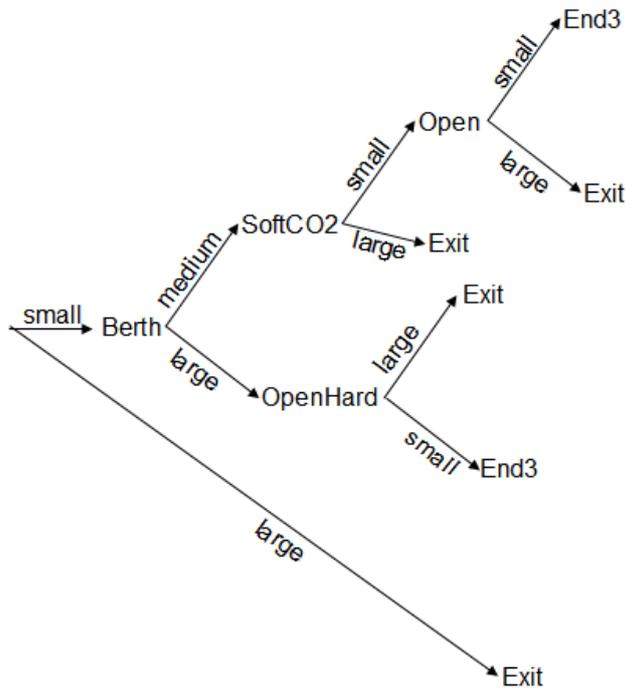

Building the DT using probabilities demands more data than SA on the one hand; however it allows us to receive more precise outcomes on the other hand. Due to the sensitivity of the data we change the numerical probabilities in the decision tree to their equivalent in words e.g. a probability that is referred to as small at the figure means $1\% < p \leq 10\%$ (see Table 12).

Table 12: probabilities that are used in the decision tree and their equivalent in words.

| Probabilities | $p > 50\%$ | $10\% < p \leq 50\%$ | $1\% < p \leq 10\%$ | $p \leq 1\%$ |
|---|---|---|---|---|
| Equivalence | large | medium | small | very small |

From the DT we can calculate the following results (Table 13), which are identical to the scenario analysis, apart from small rounding errors (due to spreadsheet calculations).

Table 13: Decision Tree results: Number of positive lorries found if CG = 0%.

| TG vs. SG | SG 0% | SG +10% | SG +20% |
|---|---|---|---|
| TG 0% | 1674 | 1841 | 2008 |
| TG 10% | 1522 | 1674 | 1826 |
| TG 20% | 1395 | 1534 | 1674 |

Using the same combined probabilities for each scenario as we have used in SA we find that the results of two approaches are almost identical in the monetary value and according to both we should adopt the same search strategy SG by 10%. One of the downsides of the DT that we have discovered during our work is its linearity. It is an easy task to build a linear DT when the data is linear, however it is more complex if the data is distributed exponentially and requires some manipulations on it. DT helps to the user to understand better the layout of the system.

*3.5 Simulation*

Besides the data already mentioned we have collected data on operation times of the activities in the shed and the berth and from ferry operation manuals. We have used the DT graphical representation as a conceptual model basis for the MCS and DES implementation. As we mentioned above the process flow representation in the simulation is equivalent to the DT layout. However, the simulation uses probabilities and frequency distributions, e.g. exponential arrival times of lorries. This randomness (e.g. slightly different number of arrivals each time) requires us to undertake several replications of each simulation scenario and to calculate the means of the replication outputs. These means we compare with the results of the DT to validate our model. We then translate the results to the monetary value and compare the outcomes to other methods.

Figure 2: The French site (AnyLogic Simulation)

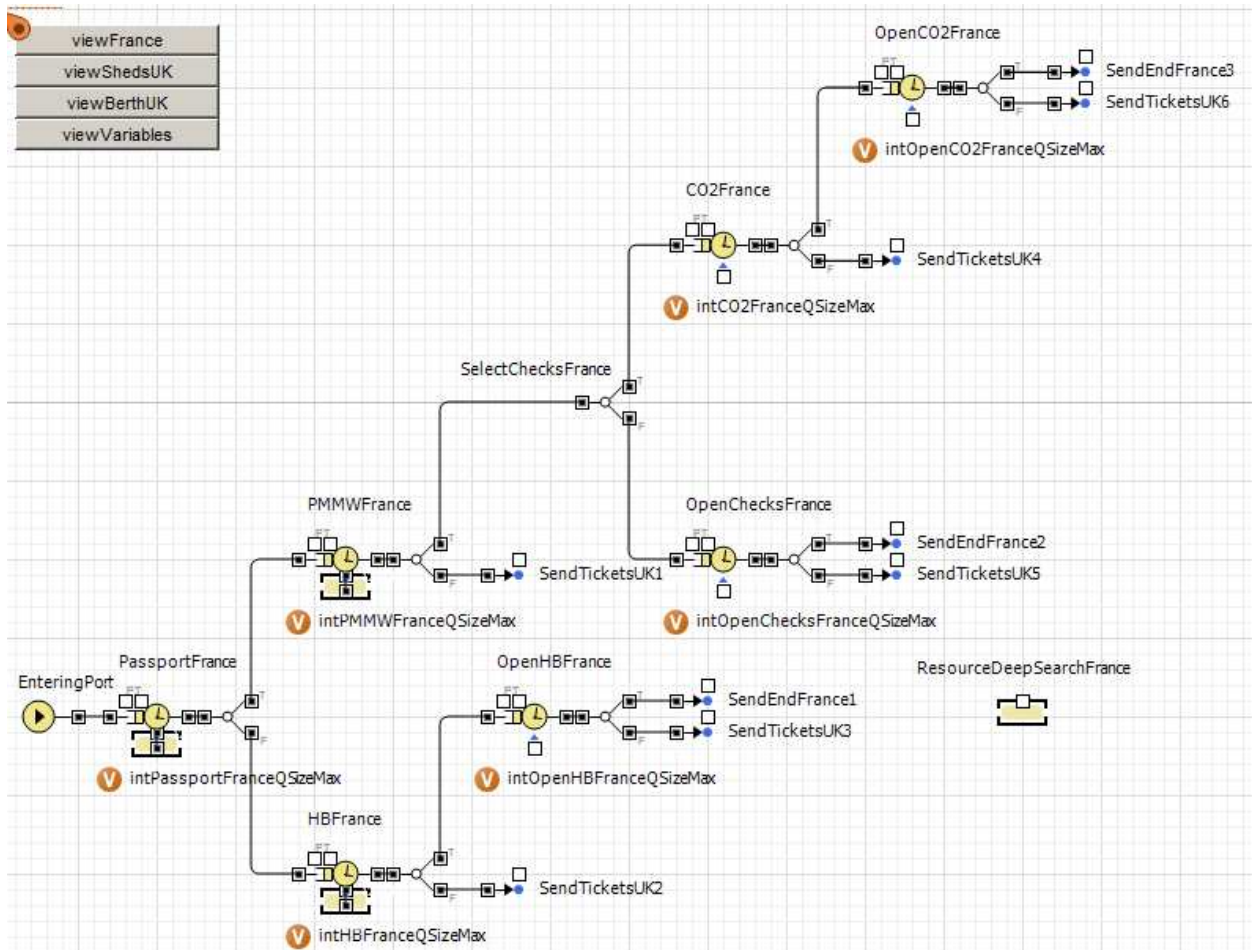

As we have mentioned already there are different types of simulation and here we use MCS and DES. We use the same simulation model for both types. The parameter set loaded in the initialisation phase of the simulation determines the type of simulation run. The first parameter set sets all delay times to zero and all queue sizes to 1,000,000. This allows emulating a MCS where all events are executed in the correct sequence but the time to execute them is zero.

The second parameter set defines triangular frequency distributions for the delay times (based on our collected case study data), sets all queue sizes to 1,000,000, and defines the resources that are available (based on our collected case study data). Queue sizes can later be restricted by defining routing rules. For both, MCS and DES, routing decision (e.g. which lorries to inspect and how to inspect them) are derived from uniform probability distributions (given probabilities are based on our collected case study data).

Having stochastic inputs means that we also have stochastic outputs. Therefore, we have to do multiple runs. We have conducted some tests to determine the number of replication required using confidence intervals (95%) following the guidelines in Robinson (2004). The test result suggests to run at least eight replications. To be on the safe side we decided to do 10 replication for each iteration of our experiments. To report our experimental results we use mean value as a point estimator and standard deviation as an estimator of the variability of the results.

### *3.6 Monte Carlo simulation (MC)*

We have run all scenarios as defined in the previous sections. The results are as follows:

Table 14: Monte Carlo Simulation results: Number of positive lorries found if CG = 0%.

| TG vs. SG | SG 0% | SG +10% | SG +20% |
|---|---|---|---|
| TG 0% | 1678.75 | 1846.25 | 2027.75 |
| TG 10% | 1531.30 | 1674.15 | 1827.50 |
| TG 20% | 1404.25 | 1540.90 | 1670.70 |

Comparing the results with the ones from DT shows the following (Table 15). We find that the differences are relatively small and can be attributed to the fact that we use a stochastic method.

Table 15: Comparing Decision Tree Results with Monte Carlo Simulation results (errors)

| TG vs. SG | SG 0% | SG +10% | SG +20% |
|---|---|---|---|
| TG 0% | -5.0 | -5.2 | -19.4 |
| TG 10% | -9.6 | -0.4 | -1.7 |
| TG 20% | -9.4 | -6.6 | 3.0 |

We then extend this MCS model to a DES model by adding elements that are linked to time (e.g. arrival rates, delays, queues) and constraints that are linked to time (e.g. queue size restrictions) and new performance measures linked to time (e.g. utilisation, time in system, max queue length) in order to demonstrate the full power of simulation as a tool in the decision making process. These modifications reflect the real world and make our simulation model more realistic than the DT. Also MCS allows us to learn more about system variability an ability of running many replications and having a range of results, can help to learn about sensitivity inside the system.

As we mentioned before we use the same combined probabilities for each scenario as we have used in SA and DT. We find that the simulation results are similar to the other two approaches. Although here, according to the simulation outcomes we should adopt the same search strategy, e.g. SG 10%. The difference in the monetary value between MCS and other two approaches is varying as result of calculating the costs according to the averages (DT and SA) vs. according to the output for the simulation.

### *3.7 Discrete event simulation 0 (DES)*

In aim to upgrade our model to DES we have added additional data that represents the real world situation. The extra data that we use is service times and resources we run our simulation again to check if this additional data makes an effect on the results. We find that the impact is very small and as a result we would choose the same strategy as before SG 10%.

The cycle times are based on data that we collected through observations and from interviews with security staff. In order to represent the variability that occurs in the real system we use different triangular distribution for each sensor types. Triangular distributions are continuous distributions bounded on both sides. In absence of a large sample of empirical data a triangular distribution is commonly used as a first approximation for the real distribution (XJTEK 2005). Every time a lorry arrives at a shed a value is drawn from a distribution (depending on the device that will be used for screening) that determines the time the lorry will spend in the shed.

Table 16: Standard Discrete Event Simulation Results: Number of positive lorries found if CG = 0%.

| TG vs. SG | SG 0% | SG +10% | SG +20% |
| --- | --- | --- | --- |
| TG 0% | 1674.50 | 1833.00 | 2013.90 |
| TG 10% | 1512.00 | 1667.20 | 1818.15 |
| TG 20% | 1387.65 | 1527.90 | 1694.20 |

To show the additional possibilities of DES we have added some additional features we want to demonstrate here: These are (1) variable arrival rates, (2) queue size restrictions at sheds UK and (3) combination of first two features.

*3.8 DES 1*

In order to be able to emulate the real arrival process of lorries in Calais we created hourly arrival rate distributions for every day of the week from a year worth of hourly arrival records that we received from the UKBA. These distributions allow us to represent the real arrival process, including quiet and peak times. In cases where this level of detail is not relevant we use an exponential distribution for modelling the arrival process and the average arrival time calculated from the data we collected as a parameter for this distribution.

After running the DES with actual arrival rates we find that also here the results are very similar to previous runs and scenarios and the appropriate strategy to adopt is SG 10%. The monetary difference is very small. The results are logical if there is no queue capacity involved and the only change is at the arrival rates.

Table 17: Discrete Event Simulation with Variable Arrival Rates results: Number of positive lorries found if CG = 0%.

| TG vs. SG | SG 0% | SG +10% | SG +20% |
|---|---|---|---|
| TG 0% | 1681.55 | 1843.00 | 2008.70 |
| TG 10% | 1519.25 | 1687.20 | 1852.85 |
| TG 20% | 1385.05 | 1534.85 | 1658.85 |

*3.9 DES 2*

At this stage we run the DES while queue capacity is set, we find that here the results are different than in previous runs and scenarios, but the appropriate strategy to adopt is still SG 10%. The monetary difference is bigger. Although, we find that these results are logical the queue capacity only has its impact on the system but this effect is not big enough to make change in the strategy.

Table 18: Discrete Event Simulation with Queue Size Restrictions results: Number of positive lorries found if CG = 0%.

| TG vs. SG | SG 0% | SG +10% | SG +20% |
|---|---|---|---|
| TG 0% | 1666.95 | 1833.30 | 1998.90 |
| TG 10% | 1507.80 | 1657.85 | 1795.35 |
| TG 20% | 1383.90 | 1526.70 | 1653.95 |

*3.10 DES 3*

As a final stage for this book chapter we have combined the three previous scenarios for DES: resources, arrival rates and queue capacity. From the results at this stage we find that the most appropriate strategy to adopt is to keep SG 0%. The reason for the different outcome lies in the combination of arrival rates and queue capacity, when the effort to search more is wasted by queue jumping during the peak times.

Table 19: Discrete Event Simulation with Variable Arrival Rates and Queue Size Restrictions results: Number of positive lorries found if CG = 0%.

| TG vs. SG | SG 0% | SG +10% | SG +20% |
|---|---|---|---|
| TG 0% | 1626.35 | 1788.10 | 1891.50 |
| TG 10% | 1500.75 | 1619.15 | 1731.25 |
| TG 20% | 1374.8 | 1492 | 1600.9 |

## 4. Conclusions

We obtained similar results for all methods we looked at, e.g. SA, DT and MC. The DES results are different from the others as we explained in the relevant sections. One could argue that the data that was used for the DT is more detailed than the one for the SA, and the data that was used for the simulation is more detailed than DT. However, as in our case, full data is often collected by the stake holders anyway, thus no "extra data" is required by simulation. Using it in simulation actually shows its strength as a tool. The summary of the results are presented in Table 20. Table 21 presents the relative cost to the lowest cost scenario for all methods: option three will be always rejected as the most expensive one and the more detailed simulation results show that the best policy is the current policy, e.g. SG=0% (Tables 20, 21).

Table 20: Overall cost comparisons of all methodologies.

| Option | Total expected costs | | | Cheapest option |
|---|---|---|---|---|
|  | 1: SG=0% | 2: SG=10% | 3: SG=20% |  |
| SA | £60,500,000 | £60,000,000 | £60,416,667 | 2 |
| DT | £60,497,446 | £60,000,000 | £60,418,795 | 2 |
| MCS | £60,335,818 | £60,058,184 | £60,461,341 | 2 |
| DES 0 | £60,797,873 | £60,250,740 | £60,350,102 | 2 |
| DES 1 | £60,881,284 | £60,017,602 | £60,406,308 | 2 |
| DES 2 | £60,714,953 | £60,166,442 | £60,857,915 | 2 |
| DES 3 | £59,817,382 | £60,116,618 | £61,624,835 | 1 |

Table 21: Relative cost comparisons of all methodologies.

| Option | Relative difference from lowest costs | | | Cheapest option |
|---|---|---|---|---|
|  | 1: SG=0% | 2: SG=10% | 3: SG=20% |  |
| SA | £500,000 | £0 | £416,667 | 2 |
| DT | £497,446 | £0 | £418,795 | 2 |
| MCS | £277,633 | £0 | £403,156 | 2 |

| | | | |
|---|---|---|---|
| DES 0 | £547,133 | £0 | £99,362 | 2 |
| DES 1 | £863,682 | £0 | £388,706 | 2 |
| DES 2 | £548,511 | £0 | £691,473 | 2 |
| DES 3 | £0 | £299,236 | £1,807,453 | 1 |

The main purpose of this chapter has been to demonstrate the difference between the tools, and illustrate their data requirement when applied to the same case study (Table 22). Also we need to remind the reader that these results are based on mixture of assumptions and real world data. In addition to our limitations we find that we cannot capture the intangible costs or benefits such as the benefits of the society as a result of a policy adopted. In our further work (Siebers et al. 2011) we will compare different DES techniques e.g. process oriented and object oriented and suggest using MCA to evaluate non monetary benefits.

Table 22: Factors to take into consideration before making decisions (SA – Scenario Analysis, DT – Decision Tree, MC – Monte Carlo Simulation, DES – Discrete Event Simulation)

| | | SA | DT | MC | DES |
|---|---|---|---|---|---|
| | Discrete / Continuous | D | D | C | C |
| Risk type | Correlated / Independent | C | I | Both | Both |
| | Sequential / Concurrent | C | S | Both | Both |
| Decision process | Strategic / Operational | S | S | S | O |
| | Broad / Detailed | B | B | B | D |
| | Difficulty | L | M | H | H |
| | Data requirements | L | L | M | H |
| Model Characteristics: (Low, Medium, High) | Tool costs | L | L | M | H |
| | Training costs | L | L | H | H |
| | Assumptions | H | M | L | L |

While dealing with continuous types of risks (i.e. illegal immigration) the simulation approach will be more suitable and it does not depend if the risk type is correlated or independent, sequential or concurrent. However, for SA and DT the situation is different. SA can be used for correlated, concurrent and discrete type of risk while DT can be used for independent, sequential and discrete types. The decision process level and the information required for the three basic approaches will be similarly applied at the strategic level and require broad information. On the other hand, DES requires detailed information and can be applied at the operational level. Total costs of the approaches (including training costs) vary from low for SA (that require just pen and paper) and DT to high for simulation.

Table 23 summarises real world data that was collected during our case study and the results and outputs that can be obtained from the model (in italics) and help in decision making (resources might be collected as part of a real world data, whereas resource utilisation can be a part of the output from the model). As we can conclude from Table 23, DES requires the largest amount and detail of data. However, it also provides many different outputs such as peak times, bottle necks, resource utilisation, system throughput and service quality (e.g. waiting times in the system). This can be so useful for the decision makers in services and other industries. This output can be very valuable for an analyst or a decision maker. It is important to clarify that the tools mentioned in Table 23 i.e. simulation or DT will not directly provide the total expected costs. The user will need to conduct some extra mathematical calculations on the outputs to obtain these costs as well as making a decision about the policy alternatives.

Table 23: Real world data the can be considered in the model and decision support output that can be collected from the model (in italics).

| Scenario Analysis | Decision Trees | Monte-Carlo Simulation | Process Oriented DES |
|---|---|---|---|
| Scenarios | Scenarios | Scenarios | Scenarios |
| *Total expected costs* | *Total expected costs* | *Total expected costs* | *Total expected costs* |
| *Positive lorries detected* | *Positive lorries detected* | *Positive lorries detected* | *Positive lorries detected* |
| | System structure | System layout | System layout |
| | Existing resources | Existing resources | Existing resources |
| | | *System variability* | *System variability* |
| | | | *Resource utilisation* |
| | | | Dynamic system constraints |
| | | | *Throughput (capacity)* |
| | | | *Waiting time distributions* |
| | | | *Bottleneck analysis* |
| | | | Dynamic system decisions |

Our future plans are to get away from discrete event process oriented approach (PO) towards discrete event object oriented approach (OO). In this approach, the intelligence (decision making mechanisms) is embedded in the objects. This way of modelling implies some new modelling challenges but it also adds to model credibility, as it allows considering more detailed decision making processes of the active entities in the real system (in our case, security officers and customers) and therefore presents a more natural way of capturing the operations of the real system. In our future research we will investigate if this approach provides additional insight into system behaviour and if the modelling method (intelligence representation style) has an impact on model outputs and/or model validity. Another aspect that we are planning to investigate is application of the same methods on a different case study (the port of Dover).

**Acknowledgements**

This project is supported by the EPSRC (EP/G004234/1) and the UK Border Agency.